\documentclass[conference]{IEEEtran}
\IEEEoverridecommandlockouts

\AtBeginDocument{%
  \providecommand\BibTeX{{%
    Bib\TeX}}}

\usepackage{algorithm}
\usepackage{algorithmic}
\usepackage{comment}
\usepackage{stfloats}
\usepackage{float} 
\usepackage{amsmath,amsfonts}
\usepackage{graphicx}
\usepackage{textcomp}
\usepackage{xcolor}
\usepackage{xspace}
\usepackage{wrapfig}
\usepackage{todonotes}
\usepackage{soul}
\usepackage{makecell}
\usepackage{subfiles}
\usepackage{subcaption}
\usepackage{enumitem}
\usepackage{fancyhdr}
\usepackage{balance}
\usepackage{caption}
\usepackage{subcaption}
\usepackage{multirow}
\usepackage{mdframed}
\usepackage[colorlinks,linkcolor=blue,citecolor=blue,urlcolor=blue,filecolor=blue]{hyperref}

\mdfdefinestyle{MyFrame}{%
    linecolor=black,
    outerlinewidth=2pt,
    roundcorner=20pt,
    innertopmargin=4pt,
    innerbottommargin=4pt,
    innerrightmargin=4pt,
    innerleftmargin=4pt,
    leftmargin = 4pt,
    rightmargin = 4pt,
    backgroundcolor=black!20,
    rightline=false,
    leftline=false,
    topline=false,
    bottomline=false
}

\newcommand{\ie}{\textit{i.e.,}\xspace}

\newcommand{\etal}{\textit{et al.}\xspace}

\newcommand{\ignore}[1]{}

\newcommand*\whitecircle[1]{%
 \protect\begin{tikzpicture}[baseline=(C.base)]                                          
 \protect\node[draw,circle,fill=white,inner sep=0.2pt](C) {\textcolor{black}{#1}};       
 \protect\end{tikzpicture}} 

\def\BibTeX{{\rm B\kern-.05em{\sc i\kern-.025em b}\kern-.08em
    T\kern-.1667em\lower.7ex\hbox{E}\kern-.125emX}}
\begin{document}

\title{Performance Characterizations and Usage Guidelines of Samsung CXL Memory Module Hybrid Prototype}

\author{
\IEEEauthorblockN{Jianping Zeng\IEEEauthorrefmark{1}
    Shuyi Pei\IEEEauthorrefmark{1} Da Zhang\IEEEauthorrefmark{1}
    Yuchen Zhou\IEEEauthorrefmark{2} Amir Beygi\IEEEauthorrefmark{1}
    Xuebin Yao\IEEEauthorrefmark{1} Ramdas Kachare\IEEEauthorrefmark{1} \\
    Tong Zhang\IEEEauthorrefmark{1} Zongwang Li\IEEEauthorrefmark{1}
    Marie Nguyen\IEEEauthorrefmark{1} Rekha Pitchumani\IEEEauthorrefmark{1}
    Yang Soek Ki\IEEEauthorrefmark{1} Changhee Jung\IEEEauthorrefmark{2}}
\IEEEauthorblockA{
    \IEEEauthorrefmark{1}Samsung Semiconductor, USA
    \\\{jp.zeng, shuyi.pei, zhang.da, a.beygi, xuebin.yao, r.kachare,\\
        t.zhang2, zongwang.li, marie.n, r.pitchumani, yangseok.ki\}@samsung.com}
\IEEEauthorblockA{
    \IEEEauthorrefmark{2}Purdue University, USA
    \\\{zhou1166, chjung\}@purdue.edu}
}

\maketitle
\thispagestyle{plain}
\pagestyle{plain}

\begin{abstract}

The growing prevalence of data-intensive workloads, such as artificial intelligence (AI), machine learning (ML), high-performance computing (HPC), in-memory databases, and real-time analytics, has exposed limitations in conventional memory technologies like DRAM. While DRAM offers low latency and high throughput, it is constrained by high costs, scalability challenges, and volatility, making it less viable for capacity-bound and persistent applications in modern datacenters.

Recently, Compute Express Link (CXL) has emerged as a promising alternative, enabling high-speed, cacheline-granular communication between CPUs and external devices. By leveraging CXL technology, NAND flash can now be used as memory expansion, offering three-fold benefits: byte-addressability, scalable capacity, and persistence at a low cost. Samsung's CXL Memory Module Hybrid (CMM-H) is the first product to deliver these benefits through a hardware-only solution, i.e., it does not incur any OS/IO overheads like conventional block devices. In particular, CMM-H integrates a DRAM cache with NAND flash in a single device to deliver near-DRAM latency.
This paper presents the first---publicly available---study for comprehensive characterizations of an FPGA-based CMM-H prototype. Through this study, we address users' concerns about whether a wide variety of applications can successfully run on a memory device backed by NAND flash medium. Additionally, based on these characterizations, we provide key insights into how to best take advantage of the CMM-H device.



\end{abstract}

\setlength{\intextsep}{5pt} 
\setlength{\columnsep}{8pt} 

\section{Introduction}


In recent years, the emergence of data-intensive workloads, e.g., artificial
intelligence (AI) \cite{winston1992artificial}, machine learning (ML)
\cite{jordan2015machine}, real-time analytics \cite{milosevic2016real}, and
high-performance computing (HPC)
\cite{vetter2015opportunities,caulfield2010understanding}, has led to
exponential increases in the demand for high-performance memory
\cite{gholami2024ai,huang2020memory,kwon2018beyond,qureshi2020tearing}.
These workloads need
efficient memory architectures that can store enormous amounts of data with low
latency and high throughput. These demands pose significant challenges in
relying solely on conventional DRAM (Dynamic Random-Access Memory), which is
attached to DDR slots, to expand memory capacity due to the following issues.



First, due to its inherent hardware structures, DRAM medium scalability is constrained by 
physical limitations, including manufacturing cost \cite{mandelman2002challenges}, 
leakage current \cite{kim2008doped}, and heat dissipation \cite{ghose2018your},
leading to limited memory size per module. Moreover, memory controllers
only support limited DDR pin counts and yield deficient memory capacity due to significant area and power overheads as well as complicated placement, routing, and packaging problems. \cite{cho2024coaxial,zhu2016package}.
All those issues render adopting DRAM less viable or at least challenging in
large-scale datacenters for the forthcoming and future capacity-bound workloads.

In response to the above issues, storage-class memory (SCM) \cite{burr2008overview}
technologies, such as STT-RAM \cite{kultursay2013evaluating,chi2016architecture},
ReRAM  \cite{akinaga2010resistive,chen2020reram,zhou2023sweepcache,
choi2023write,gan2024,choi2022compiler,choi2019cospec,liu2019hr,
liu2021fpra},
MeRAM \cite{huai2008spin,khvalkovskiy2013basic,oh2022rethinking},
PCM \cite{tyagi2007pcm,qureshi2009enhancing,kim2019ll,zeng2021replaycache},
and FeRAM \cite{mikolajick2001feram,nagel2000overview,fox2004current,
takashima2011overview}, have emerged to offer several times larger capacity than DRAM
thanks to their underlying physical materials.
For example, Intel Optane memory (PMEM) \cite{intel2023pm}---based on PCM---can
offer 2x larger
capacity than the largest DRAM DIMMs (256 GB) and is an order of magnitude larger
than typical 32 GB DRAM DIMMs used in today's datacenters \cite{shanbhag2020large}.

Although those SCM technologies offer higher density than DRAM, their capacities
and costs are still not satisfactory. This is because (1) SCM's internal hardware
structures limit rapid increase in capacity \cite{burr2014storage}
and (2) SCM is typically attached to conventional DDR slots and thus faces the same
scalability issue as DRAM.

Fortunately, the first issue of medium scalability can be addressed by NAND flash
technology \cite{compagnoni2017reviewing,goda2021recent,grupp2012bleak,
micheloni2010inside,mielke2017reliability,zhang2023excessive}. It achieves high capacity through its
dense yet simple cell design without
requiring complex peripheral circuitry. Besides, NAND flash can store multiple
bits per cell, e.g., multi-level cell (MLC) \cite{cai2015data}, triple-level
cell (TLC) \cite{mizoguchi2017data}, and quad-level cell (QLC) 
\cite{liang2019empirical}, increasing storage density significantly.
Moreover, NAND flash's simple structure makes it easier than DRAM to leverage
3D die stacking \cite{hadidi2017demystifying,black2006stacking,loh2007processor} for achieving high capacity. Consequently, while typical DRAM capacities
range from 4 GB to 32 GB per module (consumer-grade) and up to 1 TB in specialized
server-grade DRAM configurations, NAND flash capacities range from
128 GB to 8 TB (consumer-grade) and up to 100+ TB in enterprise-grade NAND storage
systems.

In addition, NAND flash consumes significantly less energy than DRAM, mitigating
heat dissipation. This is because NAND flash is nonvolatile and thus retains memory
data without a constant power supply. In contrast, due to continuous power draw,
DRAM needs to periodically refresh its cells to maintain data, which causes 
considerable energy waste. More importantly, NAND flash cells store data using a
floating gate or charge trap, requiring fewer active components such as capacitors
and transistors compared to DRAM.

Although NAND flash technology addresses the medium scalability and heat dissipation issues of DRAM, it operates on a block basis; i.e., it is not byte-addressable, as each NAND flash access involves transferring data in chunks much larger than a single byte. To allow for byte-granular access to external devices, industry efforts have recently focused on forming a consortium to develop Compute Express Link (CXL), an open standard for high-speed interconnects that enables an efficient, cacheline-granular, and scalable way to expand memory capacity \cite{sharma2022compute,das2024introduction}. Specifically, PCIe interfaces---on which CXL operates---can deliver similar bandwidth using fewer pins compared to DDR; for example, 8 PCIe 5.0 lanes offer 32 GB/s of bandwidth per direction, while DDR5-4800 requires 160 pins to achieve 38.4 GB/s \cite{cho2024coaxial}, allowing for higher memory capacity per pin and mitigating the DDR pin scalability issue.

In particular, CXL allows CPUs to access device memory, e.g., NAND flash, in a
cache-coherent manner. That is, NAND flash is now addressable by regular load\slash 
store instructions, and its data can be maintained in CPU caches, which significantly improves
the performance of cache-friendly applications. This contrasts with conventional
memory-mapped I/O \cite{habicht2024fundamental,papagiannis2020optimizing,
song2016efficient,choi2020libnvmmio} where flash memory contents cannot be
cached in CPUs. In such cases, accesses to memory regions assigned to I/O devices are non-cacheable, thus losing the opportunity to benefit from CPU caching.




Interestingly, with the help of CXL, NAND flash can now function as persistent
memory. This significantly boosts the performance of persistent applications,
e.g., in-memory databases \cite{redis7.4.1,rocksdb9.9.3}, by eliminating costly
logging to external persistent devices; please refer to Section \ref{sec:persist_mem}
for more details. The takeaway is that \textbf{\textit{CXL enables an economical way to
significantly expand main memory capacity from the GB level to the TB level without
losing persistence guarantee \cite{mutlu2013memory}.}}


\begin{figure}[h!]
    \centering
    \includegraphics[width=0.8\linewidth]{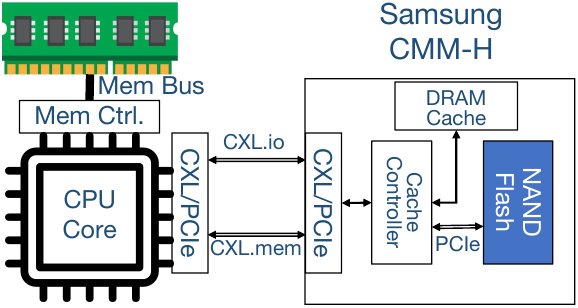}
    \caption{High-level architecture of Samsung CMM-H; assume it is connected to
    conventional CPUs, though it is technically possible to attach CMM-H to other
    accelerators (e.g., GPUs) in the future}
    \label{fig:cmmh_arch}
\end{figure}

Desiring to reap the above benefits, Samsung recently released a proof-of-concept
CXL-based memory expansion device, called the CXL Memory Module Hybrid (CMM-H)
\cite{samsung2024cmmh} for emerging capacity-bound applications. As shown in
Figure \ref{fig:cmmh_arch}, CMM-H internally employs a DRAM cache to accelerate 
access to the backend NAND flash. \textbf{\textit{Thanks to the high DRAM
cache hit rate, as shown in Section \ref{sec:xsbench}, the CMM-H device achieves
DRAM-level latency and NAND-level capacity at the same time.}}




Also, CMM-H offers persistence in a fully transparent manner. Upon impending
power loss, CMM-H flushes all volatile data in the DRAM cache to the NAND flash
by using energy stored in an energy buffer (i.e., a battery), thereby transparently
offering {\it persistence} to the upper-level software stack. Compared to conventional
NVDIMMs \cite{pmmicron} and Intel PMEM \cite{intel2023pm}, CMM-H 
can offer a much larger memory space by exposing the high-capacity backing NAND flash
to the software stack in an economical way; NVDIMMs make only DRAM space visible to
software, while PMEM is still $\approx$ 100x smaller than NAND flash.

Yet, there is no performance evaluation of CMM-H available to the public on
its basic characteristics and, in particular, on how it affects persistent applications.
\textbf{\textit{To bridge the gap, this paper, for the first time, conducts a comprehensive
performance analysis of the CMM-H device using microbenchmarks and real-world workloads from
various domains---e.g., large language models (LLMs), conventional CPU benchmark
suites, high-performance computing (HPC) program, and in-memory databases---providing
key insights on how to best use the CMM-H device.}}
In summary, this paper makes the following contributions:

\begin{itemize}
    \item We are the first to thoroughly characterize the performance of the proof-of-concept Samsung CMM-H device, including basic read and write latencies, tail latency, and bandwidth.
    \item We demonstrate how the performance of the CMM-H prototype changes
    with varying factors, e.g., thread count, memory footprint, and program behaviors.
    \item We present that, with persistence ensured, CMM-H dramatically boosts
    the performance of persistent applications.
    \item With the performance analysis from the experiments, we offer key findings
    on how to best use the CMM-H device for both volatile and persistent applications.
\end{itemize}

\section{Background}
\subsection{Compute eXpress Link (CXL)}

Compute eXpress Link (CXL) \cite{sharma2023introduction} is an open-standard,
high-speed, and low-latency interconnect designed to enhance the
performance of communication between processors, memory expanders,
accelerators, and other
peripherals. Built on the PCIe physical layer, CXL extends the capabilities of
traditional interconnects by offering coherence and memory-sharing features,
satisfying the growing demands of modern data-intensive workloads. The upshot
is that CXL's
innovative architecture enables efficient resource utilization, making it a
cornerstone of next-generation computing systems \cite{sun2023demystifying}.

For the sake of versatility and flexibility, CXL implements three protocols, {\tt CXL.io}, {\tt
CXL.mem}, and {\tt CXL.cache}. Among them, {\tt CXL.io} is technically compatible
with PCIe, i.e., {\tt CXL.io} uses the same uncacheable load-store semantics
for link initialization, 
device discovery, status reporting, virtual-to-physical
address translation, and direct memory access (DMA) \cite{das2024introduction}.
Owing to this guaranteed compatibility, existing PCIe devices are reusable for
CXL, saving vendors' existing considerable investments.

Meanwhile, CXL is engineered to provide nanosecond-level latency for memory accesses at cacheline
granularity through {\tt CXL.mem} protocol. That is, every access to CXL
memory is byte-granular and can benefit from the cache hierarchy of the host CPUs.
CXL ensures that processors
can retrieve data with lower delays, significantly speeding up access to
storage devices and ultimately improving computational
throughput. In contrast, conventional block devices operate at page granularity
(e.g., 4 KB) and thus experience significant I/O overhead---e.g., context
switching, metadata maintenance, and system calls---involved in file system
operations \cite{hsu2004performance}, due to the
mismatched granularity between page and cacheline. Even though memory-mapped
I/O, e.g., Linux mmap, can mitigate the I/O overhead, it still leads to suboptimal
performance. This is because (1) uncacheable memory-mapped I/O operations
\cite{song2016efficient} result in significant performance degradation for cache-friendly
workloads and (2) Linux mmap may generate many small and random I/Os, possibly causing scalability issues \cite{papagiannis2020optimizing}.


In addition, with the {\tt CXL.mem} protocol, CXL allows for memory pooling---a
mechanism in which multiple processors and accelerators can dynamically access
a shared pool of memory \cite{li2023pond,gouk2023memory}. This eliminates memory
silos and optimizes memory utilization, enabling systems to scale efficiently
without the bottlenecks of fixed memory assignments. For workloads requiring
vast memory capacity, e.g., in-memory databases and real-time analytics, CXL
ensures performance scalability without being limited by the finite DDR
slots.

Finally, another standout feature of CXL is its support for coherent memory
access via the {\tt CXL.cache} protocol. It enables processors and devices to share
data in a consistent and synchronized manner. This capability reduces the need
for data duplication and movement, thus decreasing latency and conserving bandwidth.
In scenarios such as GPU-accelerated computing---where accelerators rely heavily on 
accessing CPU memory---CXL's coherence can provide seamless collaboration and faster
execution of tasks.


\subsection{Samsung CMM-H and Operation Modes}

Recognizing CXL's potential, Samsung took a step further by prototyping
a proof-of-concept (PoC) memory expansion device, called the CXL Memory Module Hybrid
(CMM-H), which supports CXL v1.1. As depicted in Figure
\ref{fig:cmmh_arch}, CMM-H contains a large NAND flash, a small DRAM, and a
cache controller---based on FPGA---and comes with built-in circuitry to
ensure data persistence when enabled. The NAND flash offers greater capacity and lower cost than the
DRAM, while the DRAM functions as a hardware-managed cache and is thus invisible to
the upper-level software stack. That is, the upper-level software stack only sees the
large NAND flash. Unlike conventional caches, in the version of the PoC
device used in this paper, the DRAM cache\footnote{We
interchangeably use "DRAM cache" and "device DRAM cache" if there is no ambiguity.}
buffers {\it hot} data on a 4 KB page basis, i.e., the cache block size is 4
KB, to mitigate the long latency of accessing the NAND flash. Using page-sized
blocks also simplifies communication between the DRAM cache and the NAND flash. As such,
in contrast to conventional NAND flash and DRAM, this hybrid architecture
enables CMM-H to achieve the following 4 goals at the same time: (1) high performance owing to
DRAM cache and CXL-enabled byte addressability, (2) high capacity enabled by
NAND technology, (3) low cost due to NAND flash, and (4) nonvolatility
(persistence) provided by NAND flash.

CMM-H is exposed as a CPU-less NUMA node to the upper-level software stack,
allowing compatibility with legacy applications. In other words, there
is no need to modify existing user program as it can run directly on the
platforms equipped with CMM-H. Moreover, CMM-H has the flexibility to operate
either as volatile memory---i.e., dirty data in the DRAM cache is lost
and the data in the NAND flash are not guaranteed to be updated (i.e.,
consistent) across power failures---or as persistent memory
\cite{akinaga2010resistive,chen2020reram,hady2017platform,tyagi2007pcm,
qureshi2009enhancing,kim2019ll,sengupta2015framework,huai2008spin,
khvalkovskiy2013basic,chi2016architecture,korgaonkar2018density,
oh2022rethinking}. This feature makes CMM-H a great fit for a variety of
workload requirements.

\subsubsection{Device Architecture and Implementation}

In this paper, we evaluate an E3.L form-factor PoC device which is based on
Versal XCVM1802 FPGA\footnote{We are planning the next generation CMM-H products
based on ASIC implementation to deliver lower latency and higher bandwidth.}.
Notably, the FPGA implements a controller to manage the DRAM cache and guarantee the persistence of its data across power failures; hereafter, we refer to this controller
as a DRAM cache controller for simplicity. The CMM-H device employs
a PCIe Gen 4 ×4 NVMe SSD (TLC) with a capacity
of 1 TB as the NAND flash portion. The device connects to the host via a PCIe
Gen 4 ×8 interface and is compliant with CXL 1.1, including support for Global
Persistent Flush (GPF) \cite{das2024introduction}. As a type 3 CXL device, CMM-H
supports both the {\tt CXL.io} and the {\tt CXL.mem} protocols. 

\subsubsection{Operational Behavior}
In this study,
the evaluated CMM-H device is equipped with a 16 GB 8-way associative DRAM cache, which
uses an LRU replacement policy and an MRU insertion policy. To quickly retrieve tags
for cache accesses, an on-chip memory is dedicated to storing tags, as in prior
techniques \cite{huang2014atcache,dong2010simple}. However, storing tags in DRAM either incurs
significant performance degradation due to doubled DRAM accesses or requires modifications on
DRAM architectures for high performance \cite{loh2011efficiently,qureshi2012fundamental,
hong2024bandwidth}. Because of this, the tag lookup and data probing in the DRAM
cache happen sequentially.

To be specific, when the last-level cache (LLC)
issues a read request (64 B granularity), the DRAM cache controller performs a
tag lookup to check if the corresponding data block already resides in the DRAM
cache. On a cache hit, the corresponding 64-byte data is returned immediately to
the LLC. Otherwise, a 4 KB page is fetched from the NAND flash via NVMe commands
and then placed into an available cache set, allowing the desired 64-byte
data to be returned to the LLC later. Of course, the cache controller should
first evict a cache block to the NAND flash---which is governed by LRU
replacement policy---if the corresponding set in the DRAM cache is full.

Similarly, when the LLC issues a write request, the cache controller first
checks whether the target data block resides in the DRAM cache by consulting
the tag store. If so, the incoming data is merged into the target block,
avoiding access the backing NAND flash---which is several orders of
magnitude slower than DRAM---and allowing program to benefit from the fast DRAM
cache. Otherwise, the controller first fetches the corresponding block from the NAND
flash and then fills the block in the DRAM cache. Of course, before fetching,
the DRAM cache controller evicts a block to the NAND flash from the DRAM cache if the
corresponding set in the cache is full.

\subsubsection{Persistence Support}
Beyond its use as large volatile memory, CMM-H can also function as persistent
memory by enabling some specific configurations in the BIOS settings. In this mode,
all (dirty) data blocks in CMM-H's DRAM cache are ensured to be persistent across
abrupt power loss, which is enabled by CXL Global Persistent Flush (GPF) 
\cite{das2024introduction}. That is, upon a detected power outage, the DRAM cache
controller receives a signal from the host to flush the dirty blocks from the
DRAM cache all the way down to the NAND flash by consuming the energy in a battery.
In addition to the reactive dirty block flushing,
CMM-H also supports proactive data flushing as defined by the GPF protocol. For example, upon receiving a GPF
forceful flushing signal, the DRAM cache controller writes back all dirty blocks
in the DRAM cache to the NAND flash, maintaining consistent memory states.


With the synergistic combination of DRAM cache and NAND flash, CMM-H achieves
flexibility in operating as either volatile or persistent memory while delivering
high performance and high capacity on the cheap. This enables users to optimize their applications
for performance, capacity, and persistence under diverse workload scenarios. The
following sections detail our experimental evaluation and analysis, highlighting
the performance characteristics of the CMM-H prototype and offering guidelines
for the efficient use of the CMM-H device.

\vspace{-5pt}
\subsection{Idempotence for Crash Consistency}\label{sec:idem}

\begin{figure}[h!]
    \centering
    \includegraphics[width=1.0\linewidth]{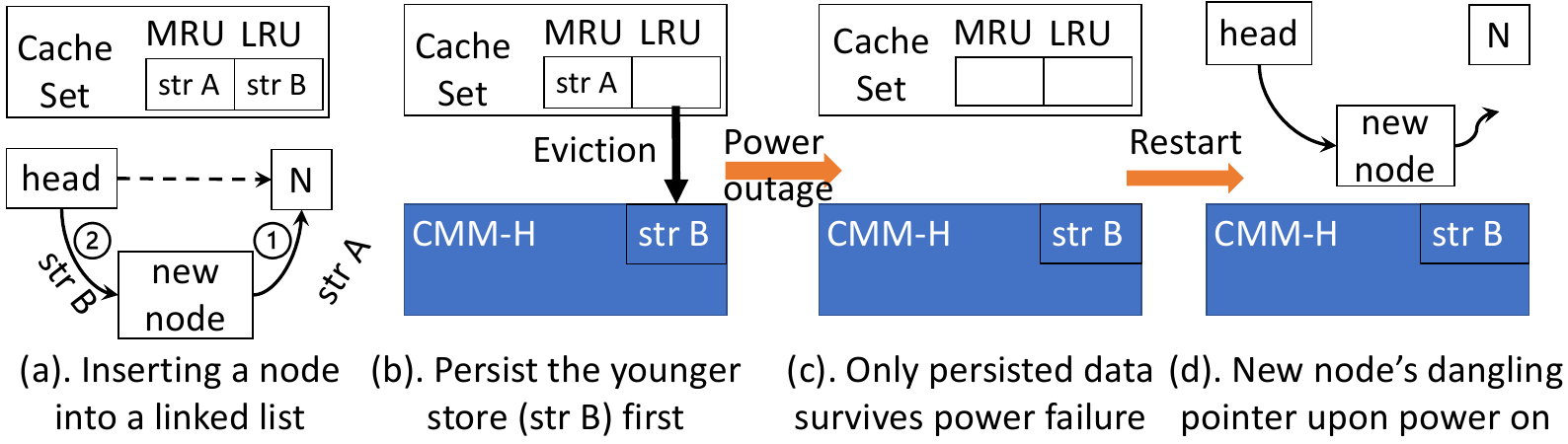}
    \caption{Inconsistent program states for singly-linked list insertion across
    power failure; CMM-H device here functions as persistent memory}
    \label{fig:linkedlist}
\end{figure}

Despite CMM-H's ability to preserve data across power failures, it
does not necessarily mean that program states are automatically crash-consistent 
\cite{bhattacharyya2022nvmr,zeng2023persistent,jeong2022capri,ren2015thynvm,
zhao2013kiln,jeong2021pmem,thesis2024zeng,zhou2024lightwsp,kim2020compiler,
zeng2021turnpike,zeng2024soft,huang2023rtailor,fang2025ipex,fang2025rethinking}.
For instance, as shown in Figure \ref{fig:linkedlist} (a), suppose a user attempts
to insert an already-created new node to the beginning of a singly-linked list.
This process
involves two steps. First, set the {\tt next} pointer of the new node to the
address of the list's first node {\tt N} (\whitecircle{1}).  Second, set the
{\tt next} pointer of the head node to the address of the new node
(\whitecircle{2}). Now, assume a scenario, depicted in Figure \ref{fig:linkedlist}
(b), where the second store ({\tt str B}) has already been evicted from the LLC
and thereby persists in CMM-H---assuming it operates as persistent memory---before
the first store ({\tt str A}) does so. If
a power outage occurs here, all volatile data---including {\tt str A}---in the
cache are lost, while only the data in the CMM-H are preserved as shown in
Figure \ref{fig:linkedlist} (c). Once power is restored, the {\tt next} pointer
of the new node becomes dangling, as illustrated in Figure \ref{fig:linkedlist}
(d), resulting in inconsistent program states.

Figure \ref{fig:linkedlist} shows that the root cause of crash
inconsistency is the loss of all volatile cache contents upon power failure. To address
this, many techniques, e.g., extended asynchronous DRAM refresh (eADR)
\cite{intel2021eadr} and CXL Global Persistent Flush (GPF) \cite{sharma2023introduction},
were proposed to include the entire cache hierarchy in persistent domain; assuming
the main memory (i.e., CMM-H) is already in persistent domain. That way, a store
becomes persisted as soon as its data is merged into the L1D cache. The beauty of
this approach is that programmers can circumvent using expensive persist
barriers---e.g., {\tt clwb} and {\tt sfence} in x86---which cause frequent
pipeline stalls and result in significant performance loss.

Unfortunately, CXL GPF or eADR alone cannot guarantee crash consistency for user program.
This is because volatile registers in the cores are still suffering from losing
their data upon power outages, preventing applications from recovering. To
address this issue, Zeng et al. \cite{zeng2024compiler} leverage compiler techniques to store
(checkpoint) registers to persistent memory, which unifies register consistency with
memory consistency. That way, each checkpoint location serves as a recovery
point from which a power-interrupted program resumes its execution, forming a
series of recoverable intervals and achieving interval-level crash consistency.

Nevertheless, checkpointing registers too often---i.e., short
checkpointing intervals---incurs high performance overhead due to frequent
write traffic towards memory; these checkpoints are essentially stores but exhibit
poor locality, as the data is only accessed during recovery. On
the other hand,
checkpointing registers too rarely---i.e., long checkpointing intervals---incurs high recovery cost; once power is restored, the recovery runtime has to roll
back the program execution to a distant previous checkpoint to recover the interrupted
program, as all the register updates since the last checkpoint are lost. This renders the instruction execution since the last checkpoint wasted.

In addition, it is not always safe to re-execute program from the last
checkpoint due to memory write-after-read (WAR) dependence---a.k.a
antidependence \cite{aho2007compilers}---which is prevalent in program code.
That is, at recovery time, a checkpoint interval with memory WAR dependences
may use updated memory data as input for its re-execution and thereby generates
incorrect program output, resulting in failed program recovery. Prior approaches
\cite{wan2016empirical,ogleari2018steal,joshi2017atom,arulraj2016write,
jeong2018efficient,haubenschild2020rethinking,jeong2020unbounded,
izraelevitz2016failure} to this problem have to resort to logging techniques
which amplify memory writes and rely on costly persist barriers. This is,
it logs memory writes followed by persist barriers in case of power failure.
For example, undo logging saves the old data
of a store to a dedicated logging area in persistent memory before
the store's memory location is updated with the new data. When a power outage occurs,
their recovery runtime reverts the memory update using the undo log, ensuring
consistent memory states.

To address the above issue while maintaining a suitable interval length for low
recovery overhead, Zeng et al. propose using 
an idempotent-processing-based compiler \cite{liu2018ido,de2011idempotent,
de2012static,de2013idempotent,kim2020compiler} to divide program into a series of idempotent
regions (intervals)---which do not have memory WAR dependences. Because of this, these regions can be re-executed arbitrarily while generating the same output. Moreover,
idempotent processing forms regions with moderate sizes, i.e.,
$\approx$ 40 instructions on average in idempotent regions \cite{zeng2024compiler},
which enables a sweet spot
between lower register checkpointing overhead and fast failure recovery. In
particular, to further reduce checkpointing overhead, the compiler performs liveness
analysis \cite{aho2007compilers} to figure out each region's live-out registers
to be checkpointed. That way, upon power failure, an interrupted program
can resume execution from the beginning of the power-interrupted region after
reloading the region's live-in registers from main memory.  Consequently,
CMM-H with idempotent processing together ensures crash consistency for user
applications.

\vspace{-5pt}
\section{Evaluation Methodology}

\subsection{System and Device Configurations}

\noindent{\textbf{System Configurations: }}
We conducted our experiments on an x86-64 dual-socket server (see Table 
\ref{tab:sys_config}). Each socket has 48 AMD EPYC 9454 cores with hardware
hyper-threading and frequency boost disabled. Each core is equipped with 32 kB L1I/L1D
caches and a 1 MB L2 cache, and each CPU has a 256 MB L3 cache. The total
DRAM main memory size is 512 GB with 256 GB per socket. We install Ubuntu 22.04.4 LTS with Linux kernel 6.8.0-49 on the server. All evaluated applications are compiled using the default GCC 11.4.0 compiler with default compilation flags unless noted otherwise.

\noindent{\textbf{Samsung CMM-H Configurations: }}
As shown in Table \ref{tab:cmmh_config},
we configure the 1 TB Samsung CMM-H as a CPU-less NUMA node so that it can be accessed
by the cores without any changes to program source code. Within the CMM-H, the 1 TB NAND flash is connected to the DRAM cache through PCIe Gen4 x4. Here, the DRAM cache is 16 GB and is managed by an FPGA board working as a DRAM cache controller.

\begin{table}[h!]
    \centering
    \caption{Host System Configurations}
    \resizebox{1.0\linewidth}{!}{
    \begin{tabular}{c||c}
        \hline
         Component & Description \\
         \hline
         OS (Kernel) & Ubuntu 22.0.4 LTS (Linux kernel 6.8.0-49) \\
         \hline
         Compiler & GCC 11.4.0 \\
         \hline
         CPU & \makecell{2x AMD EPYC 9454 CPUs @2.75 GHz, \\
         48 cores and 32 kB L1I/L1D caches, \\ 1 MB L2 cache, and 256 MB LLC per CPU \\
         Hyper-Threading disabled} \\
         \hline
         Memory & \makecell{Socket 0: 8x DDR5-4800, 300 GB/s max. bandwidth \\
         Socket 1: 8x DDR5-4800, 300 GB/s max. bandwidth \\
         Socket 2: Samsung CMM-H} \\
         \hline
         Disk & 960 GB NVMe Samsung M.2 SSD \\
         \hline
         \makecell{Host and Disk\\Connection} & PCIe Gen4 x8 \\
         \hline
    \end{tabular}
    }
    \label{tab:sys_config}
\end{table}

\begin{table}[h!]
    \centering
    \caption{CMM-H Configurations}
    \resizebox{1.0\linewidth}{!}{
    \begin{tabular}{c||c}
         \hline
         Component & Description \\
         \hline
         CXL & v1.1, CXL.io/CXL.mem (Type 3) \\
         \hline
         Device DRAM Cache & \makecell{8-way, 4 KB cacheline, LRU replacement \\ policy, MRU insertion policy, \\ writeback, 16 GB DDR4-2666} \\
         \hline
         NAND Flash & 1 TB Samsung NAND SSD (TLC) \cite{pm3a_perf} \\
         \hline
         \makecell{Device Cache-NAND \\ Flash Connection} & PCIe Gen4 x4 \\
         \hline
         Total Power & \makecell{40 W, including FPGA-based controller, \\ NAND flash, and DRAM Cache} \\
         \hline
    \end{tabular}
    }
    \label{tab:cmmh_config}
    \vspace{-5pt}
\end{table}

\subsection{Experimental Configurations}

As depicted in Table \ref{tab:sys_config}, Samsung's CMM-H memory behaves like a
CPU-less NUMA node and therefore can directly replace DRAM DIMMs as main memory
without any software changes. In our study, we treat local DRAM memory
(DDR5-L)---it is placed on the same socket as the core where the evaluated benchmarks
run---as baseline and compare it with other configurations, e.g., remote
DRAM memory (DDR5-R) and CMM-H. \textbf{\textit{Notably, we do not conduct experiments
on Intel Optane memory as it is not supported on our AMD CPUs-based testbed.}} To
evaluate program performance on the servers equipped with multiple memory devices, we use
a Linux command {\tt numactl} to force page allocation accordingly and to bind user
applications to a fixed CPU core.

\subsection{Microbenchmarks}
\label{sec:microbench}
Thanks to the CXL-enabled byte addressability, the large-capacity CMM-H device now
can be directly accessed with regular load\slash store instructions while alleviating any
software changes. To demonstrate the impact of such a great feature on application 
performance, we conduct a thorough performance characterization of the CMM-H prototype.
We employ two distinct microbenchmarks. The first is Intel Memory Latency Checker
(MLC) v3.11b \cite{intel2024mlc}, which performs pointer-chasing in a (1800 MB) memory
region---it is larger than the LLC capacity---to measure memory latency. In addition,
we follow prior approaches \cite{sun2023demystifying,yang2020empirical} to characterize
different memory devices---i.e., DDR5 and CMM-H---for four
types of instructions: temporal load ({\tt ld}), non-temporal load ({\tt nt-ld}),
temporal store ({\tt st}), and non-temporal store ({\tt nt-st}). This
provides a comprehensive performance characterization of CMM-H, allowing for
capturing key performance metrics that are critical for system designers and
programmers.

\subsection{Benchmarks}

\noindent{\textbf{Volatile applications: }} First, we run SPEC CPU2017
\cite{bucek2018spec,limaye2018workload,singh2019memory}, which is a classic performance
evaluation tool for CPUs, memory systems, and compilers, to demonstrate how CMM-H
affects program performance by replacing DRAM with CMM-H. To stress the memory system,
we run the CPU2017 applications with reference data input and collect their execution
time. Second, we execute LLaMA 3.c \cite{llama3.c}, a C version of popular large
language models (LLMs), with varying thread count from 1 to 32 to showcase how it
performs on CMM-H. Note that we feed the LLaMA 3 with llama3.2 1B  model 
\cite{llama3.2_1b_model} as data input---whose memory footprint is 4740 MB---to
isolate caching effect as much as possible and thus to stress the main memory. To 
mitigate the varying throughout of each run, we execute the LLaMA 3 program for each
thread count 5 times and average their throughputs (tokens per second). Last, we
run XSBench \cite{tramm2014xsbench}, a key computational kernel of the Monte Carlo
neutron transport algorithm, to characterize the performance HPC program on the
CMM-H. As XSBench provides a flexible way to adjust problem sizes and hence its memory
footprint, we also use XSBench with varying problem sizes to characterize how the
DRAM cache within the CMM-H prototype affects program performance; please see Section
\ref{sec:xsbench} for more details.

\noindent{\textbf{In-memory databases (persistent applications): }}
As persistent memory, CMM-H obviates
the need to periodically log program states to external slow storage devices.
To demonstrate how such a feature boosts program performance, we select two typical
in-memory databases, RocksDB 9.9.3 \cite{rocksdb9.9.3} and Redis 7.4.1 \cite{redis7.4.1}.
First, we run single-threaded {\tt db\_bench} on a specified memory device (e.g., DDR5-L) with varying
operations (e.g., {\tt fillseq} and {\tt fillrandom}). For each run, we insert 1 million
pair of key/values with their sizes set to 16 bytes and 100 bytes, respectively. We set 
block size and write buffer size to 4 kB and 64 MB by default, respectively.
Likewise, we run Redis server on a specified memory device with a single thread and use
{\tt redis-benchmark} shipped with Redis as a client to benchmark the Redis server's
throughput (requests
per second). To stress the memory, for each run, the \texttt{redis-benchmark}---running on
the DDR5-L and local core---initiates 200000 requests with a key size of 10 kB and a
random key length of 100000.

\section{Basic Latency and Bandwidth Characterization}
In this section, we first characterize the load and store latencies of
random accesses to the CMM-H device using the microbenchmarks described in Section
\ref{sec:microbench}. We then evaluate CMM-H's tail latency and bandwidth.
With these experiments in mind, we draw key takeaways on programming for
the CMM-H device.

\subsection{Load and Store Latencies}

\begin{figure}[h!]
    \centering
    \includegraphics[width=1\linewidth]{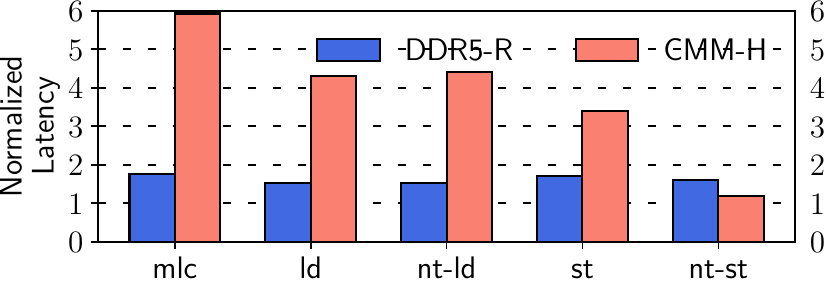}
    \caption{Normalized random access latencies of DDR5-R and CMM-H to those
    of DDR5-L (local DRAM); lower is better}
    \label{fig:micro_rand_lat}
\end{figure}

Figure \ref{fig:micro_rand_lat} reveals the normalized load and store
latencies of remote DDR5 memory (DDR5-R) and CMM-H, relative to those of local DDR5
memory (DDR5-L). The first group of bars shows the
latency measured using Intel MLC, which performs
pointer-chasing operations in a memory region of 1800 MB; it is
larger than the LLC size (i.e., 256MB per CPU). Intel MLC issues memory reads back-to-back,
which essentially measures the latency of serialized memory reads.

The remaining four groups of Figure \ref{fig:micro_rand_lat}
 show the latencies of four types of instructions: temporal load
({\tt ld}), non-temporal load ({\tt nt-ld}), temporal store ({\tt st}),
and non-temporal store ({\tt nt-st}).
Similar to prior approaches \cite{yang2020empirical,
sun2023demystifying}, we measure the latency of each instruction type by issuing 16 64-byte memory accesses to the main memory.
To avoid the impact of data locality as much as possible, each of these
memory operations accesses a random cacheable location within the main
memory. To ensure no interference with other instructions, we flush the
entire core pipeline followed by a memory fence before issuing these
memory instructions; they are also followed by a memory fence to ensure
their completions. It is
worth noting that these 16 memory accesses could be executed in parallel, 
therefore effectively measuring
the latency of random parallel memory accesses for each instruction type on a given memory device. To accurately estimate the execution time of these memory
operations, we use inline assembly code (i.e., {\tt rdtsc} instruction
in x86\_64) to read the timestamp counter twice: once before
issuing these memory operations and once after they complete.
We then subtract
the two readings to calculate the execution time of these 16 memory
accesses. We repeat the procedure of pipeline flushing, executing 16
memory accesses, reading the timestamp counter 10000 times, and calculating the
median number.

\begin{table}[h!]
    \centering
    \caption{Median Latency (ns) of three Memory Devices}
    \resizebox{1.0\linewidth}{!}{
    \begin{tabular}{c||c|c|c|c|c}
         \hline
         Memory Device & Intel MLC & ld & nt-ld & st & nt-st \\
         \hline
         DDR5-L & 122.9 & 13.1 & 13.1 & 33.5 & 13.5 \\
         \hline
         DDR5-R & 216 & 20.0 & 20.0 & 56.7 & 21.8 \\
         \hline
         CMM-H & 728.9 & 56.7 & 57.5 & 114.9 & 16.0 \\
         \hline
    \end{tabular}
    }
    \label{tab:absolute_lat}
\end{table}

Figure \ref{fig:micro_rand_lat} shows that CMM-H is slower than both DDR5-L and
DDR5-R for MLC, {\tt ld}, {\tt nt-ld}, and {\tt st}. This is expected because of
the additional latency introduced by signal propagation over the PCIe link and CMM-H's DRAM
cache controller processing. Note that the latencies measured by Intel MLC are
9x-12x higher than those measured by {\tt ld} for all three memory devices.
As presented in Table \ref{tab:absolute_lat}, for DDR5-L, Intel MLC reports a latency
of 122.9 ns, while {\tt ld} yields only 13.3 ns. This matches what
prior work observed \cite{sun2023demystifying}: Intel MLC issues read operations
back-to-back and hence cannot make use of the full-duplex PCIe link \cite{sharma2022compute}.
In contrast, {\tt ld} does so by issuing 16 memory accesses in parallel,
thereby overlapping their memory access latencies to some degree.

\begin{mdframed}[style=MyFrame]
    {\it Key takeaways: Due to the FPGA-based DRAM cache controller and
    the unmodified NAND flash architecture designed for PCIe Gen4x4 SSDs,
    CMM-H incurs 1.2x to 5.9x higher latency compared to DDR5-L. This
    indicates that latency-sensitive workloads would suffer significant
    performance degradation if they rely on CMM-H as their sole
    main memory, as confirmed in Section \ref{sec:volatile_mem}.}
\end{mdframed}


As the first two groups of bars contrast in Figure \ref{fig:micro_rand_lat},
CMM-H experiences a higher reduction in latency as the measurement switches
from Intel MLC to {\tt ld}; this is also observed in the first two columns
of Table \ref{tab:absolute_lat}. The reason, as discussed in prior work 
\cite{sun2023demystifying},  is that the sequential memory accesses issued by Intel MLC
result in a burst of cache-coherence related lookups in the remote CPU, which
unfortunately introduces extra delays for DDR5-R accesses; each remote memory access needs to go through the remote CPU's internal cache
hierarchy to search for the corresponding cache block, and even worse the
resulting latency goes up as the number of cores rises. On the contrary, 
accesses to CMM-H do not need to go through such time-consuming lookups
due to the lack of inter-core connections and cache hierarchies within the CMM-H
prototype.

The same explanation above is also applicable to the contrast between the
{\tt ld} bars and the {\tt st} bars. As illustrated in the table, compared
to DDR5-L, {\tt st} causes a higher increase (i.e., 1.70x) in latency than
{\tt ld} (i.e., 1.53x) for DDR5-R. This contrasts with what is observed
for CMM-H where {\tt st} gives a 3.4x higher latency, whereas {\tt ld} leads
a 4.3x higher latency. This is because the execution of {\tt st} instructions
need to first implicitly perform memory read accesses (i.e., {\tt ld})---
bring the corresponding cachelines from main memory to the L1D cache---for
write-allocate writeback caches. Consequently, stores experience longer execution delays than loads on DDR5-R. 
On the other hand, the execution of
{\tt st} on CMM-H avoids the costly inter-socket cache-coherence lookups which
are necessary for DDR5-R, and this latency saving outweighs the extra delay
caused by the implicit memory read accesses.


Interestingly, as shown in the last two columns of Table \ref{tab:absolute_lat},
{\tt nt-st} is faster than {\tt st} on all the three memory devices. This is
because {\tt nt-st} directly heads to main memory without going through multiple
levels of cache hierarchy, whereas {\tt st} implicitly requires a memory read
access---doubling its execution latency of {\tt st}---as discussed before.
Moreover, {\tt nt-st} performs even faster on CMM-H than on DDR5-R, due to
absence of costly inter-socket cache-coherence lookups on
CMM-H, which further lowers the execution latency of {\tt nt-st} on CMM-H.


\begin{mdframed}[style=MyFrame]
    {\it Key takeaways: {\tt nt-st} performs significantly (i.e., 7x) better than
    {\tt st} on CMM-H. This suggests that a potential way to boost the performance
    of applications with random (cache-unfriendly) writes is to replace temporal
    stores with non-temporal stores.}
\end{mdframed}

\subsection{Tail Latency}\label{sec:tail_latency}


In addition to basic load and store latencies, tail latency is also a critical
factor that affects the worst-case performance of computing systems, as memory accesses may, in the
worst case, be served from the slow backing NAND flash. For this reason,
we follow the same methodology as prior work \cite{yang2020empirical} to measure
the 99.99th, 99.999th, and maximal latencies across varying memory region sizes.
For each memory region size, a latency measurement program first flushes all the
cachelines within the memory region and then clears the core pipeline by issuing a
sufficient number of {\tt nop} instructions. After that, the program generates 16
64-byte memory read accesses to {\it random locations} within the region. For
each region size, we run the measurement program 20 million times. 

\begin{figure}[h!]
    \centering
    \includegraphics[width=0.8\linewidth]{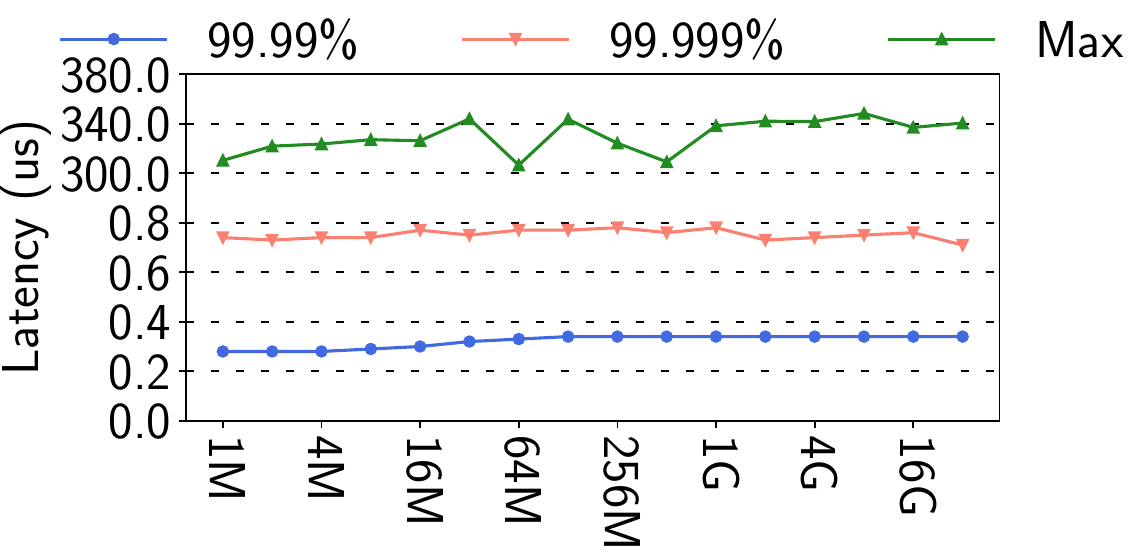}
    \caption{Tail latency in microseconds of reads for DDR5-L across
    memory region sizes}
    \label{fig:tail_lat_ddr5l}
\end{figure}

\begin{figure}[h!]
    \centering
    \includegraphics[width=0.8\linewidth]{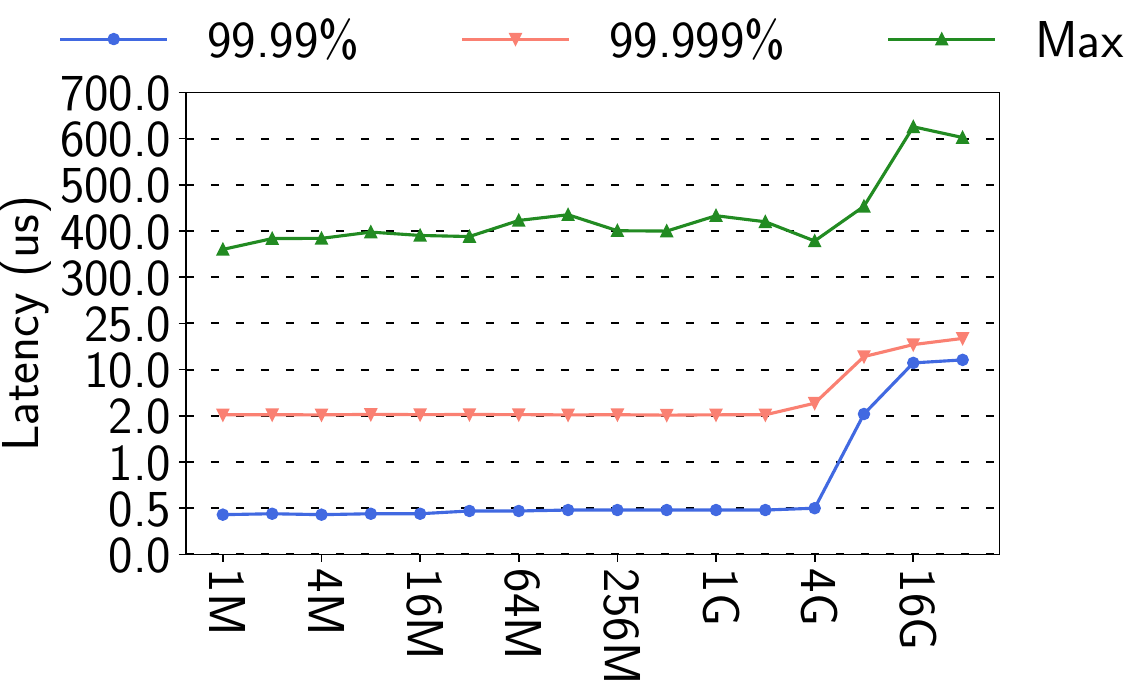}
    \caption{Tail latency in microseconds of reads for CMM-H across
    memory region sizes}
    \label{fig:tail_lat_cmmh}
\end{figure}

Figure \ref{fig:tail_lat_ddr5l} and \ref{fig:tail_lat_cmmh} show the tail latencies
of DDR5-L and CMM-H, respectively, across varying memory region sizes---which
the 16 memory reads access---from 1 MB to 32 GB. As depicted in Figure 
\ref{fig:tail_lat_ddr5l}, the 99.99th and 99.999th latencies of DDR5-L are $\approx$
300 ns and $\approx$ 750 ns, respectively, while the maximal latency of DDR5-L
fluctuates between 300 $\mu s$ and 400 $\mu s$.

%

Figure \ref{fig:tail_lat_cmmh} shows that for CMM-H, the 99.99th, 99.999th,
and maximal latencies are $\approx$ 450 ns, $\approx$ 1.8 $\mu s$, and
$\approx$ 400 $\mu s$, respectively, which are higher than those of DDR5-L;
its tail latency follows the same trend as Zhang \etal observed \cite{zhang2025skybyte}.
This is not surprising as CMM-H suffers from propagation delays over
PCIe and processing overhead from the PoC's FPGA-based DRAM cache controller.
As expected, unlike DDR5-L,
all three latencies of CMM-H rise quickly as the memory region size
exceeds 4 GB. We suspect this is because the DRAM cache experiences
significant conflict misses \cite{hill1989evaluating}, 
exposing the long latency of the backend NAND flash to upper-level program
due to the deficient DRAM cache associativity (i.e., 8).
These latencies even reach $\approx$ 15 $\mu s$, $\approx$ 20 $\mu s$, and
$\approx$ 600 $\mu s$, respectively, when the region size rises up to 32 GB;
please refer to Section \ref{sec:xsbench} for detailed analysis.

\begin{mdframed}[style=MyFrame]
    \textit{Key takeaways: When a program’s memory footprint fits within CMM-H’s
    device DRAM cache, the 99.99th, 99.999th, and maximum latencies of CMM-H
    are comparable to those of DDR5-L. However, these latencies for CMM-H
    increase significantly once the memory footprint exceeds the device DRAM
    cache capacity.}
\end{mdframed}

\subsection{Bandwidth}


Memory bandwidth is another critical factor for performance; prior approaches
\cite{zhang2021quantifying,microsoft_overclocking} have shown that higher memory bandwidth significantly
improves the performance of bandwidth-sensitive program (e.g., HPC
and ML workloads) since they typically  run multiple threads on
many cores to deliver high throughput. In this section, we measure the 
bandwidth of three instruction types ({\tt ld}, {\tt st}, and {\tt nt-st}) 
following the methodology of prior studies \cite{sun2023demystifying,yang2020empirical}. 
We exclude {\tt nt-ld} from this experiment since its behavior is similar to
that of {\tt ld}. For sensitivity analysis, we vary the number of threads
from 1 to 32. Specifically, we launch a given number of threads that execute
one of the three instruction types to access 512 cachelines and sum the
bandwidths across all threads. To maintain accuracy, we run the measurement
program 10 times for each given thread count.

\begin{figure}[h!]
    \centering
    \includegraphics[width=0.8\linewidth]{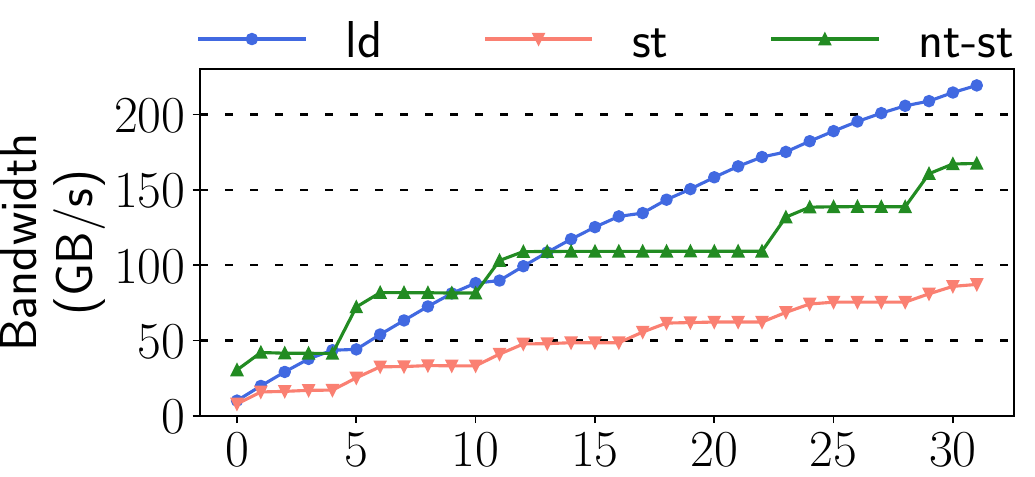}
    \caption{Bandwidth of DDR5-L across 3 kinds of instructions with
            varying threads from 1 to 32; 64 bytes access size}
    \label{fig:ddr5l_bw}
\end{figure}


\begin{figure}[h!]
    \centering
    \includegraphics[width=0.8\linewidth]{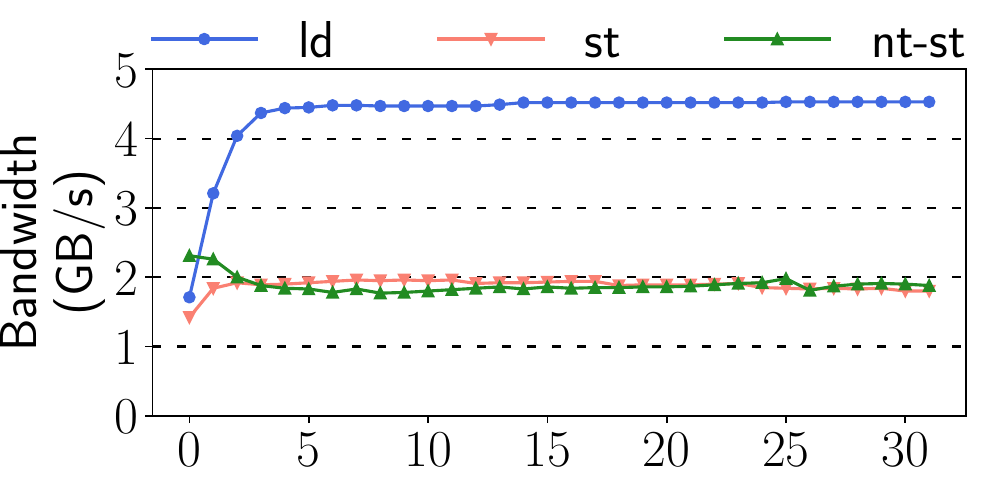}
    \caption{Bandwidth of CMM-H across 3 kinds of instructions with
            varying threads from 1 to 32; 64 bytes access size}
    \label{fig:cmmh_bw}
\end{figure}

Figure \ref{fig:ddr5l_bw} shows the bandwidth (in GB per
second) of the three instruction types for DDR5-L. We conclude that (1)
{\tt ld} delivers higher bandwidth than the other two types
of instructions, and (2) the bandwidths of these three instruction types scale well with increasing thread counts on DDR5.

Figure \ref{fig:cmmh_bw} shows the bandwidth of the three instruction types for CMM-H. 
The maximal bandwidths (i.e., 4.5GB/s) of all three instruction types running on CMM-H are significantly lower than
those (i.e., 219GB/s) on DDR5-L. Moreover, the
bandwidth of CMM-H for {\tt ld} and {\tt st} saturates at 4 and 2 threads,
respectively. Even worse, the bandwidth of {\tt nt-st} drops 
when the thread count exceeds four. 
These low bandwidth numbers suggest that the CMM-H's internal DRAM cache controller
and untailored NAND flash architecture significantly limits its performance.

\begin{mdframed}[style=MyFrame]
    \textit{Key takeaways: Compared to DDR5, the FPGA-based CMM-H has limited bandwidths,
    particularly for write operations. This limitation stems from the underlying unmodified
    NAND flash and could be overcome by employing a tailored NAND flash architecture
    that fully exploits available internal parallelism.}
\end{mdframed}

\section{Characterizing CMM-H as Volatile Memory}\label{sec:volatile_mem}

With the support for CXL-compliant interfaces, CMM-H can operate as a byte-addressable
remote memory expansion. However, CMM-H with a hybrid design combining a DRAM cache and
NAND flash cannot serve as the sole main memory for many applications. To understand
whether a wide variety of applications can successfully run on CMM-H and whether doing so meets their economic needs, in this section, we evaluate several typical workloads,
e.g., SPEC CPU2017 \cite{bucek2018spec}, LLaMA 3 \cite{llama3.c}, and XSBench
\cite{tramm2014xsbench}, on CMM-H; For comparison, we also evaluate the workloads on DDR5-L and DDR5-R.


\subsection{CPU2017}

\begin{figure}[!ht]
    \centering
    \includegraphics[width=1\linewidth]{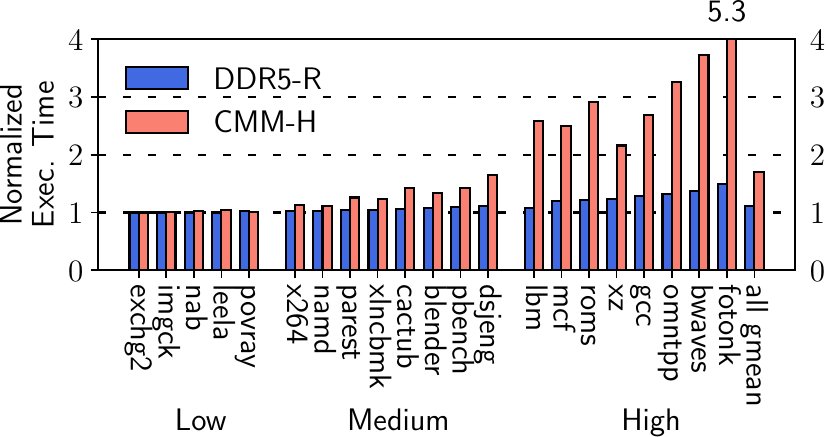}
    \caption{Normalized execution times of CPU2017 applications on DDR5-R and
    CMM-H to those on DDR5-L (baseline); lower is better; the applications are
    grouped into three kinds according to their memory pressure}
    \label{fig:cpu2017_slowdown}
\end{figure}

Figure \ref{fig:cpu2017_slowdown} shows the normalized execution times of CPU2017
applications on DDR5-R and CMM-H,  respectively, relative to those on DDR5-L; we evaluate
all applications from CPU2017 except for {\tt wrf} and {\tt cam4} due to compilation
issues. As shown in the figure, CPU2017 gets slowed down by 12\% and 70\% on average
when running on DDR5-R and CMM-H, respectively, compared to DDR5-L.

To better present the evaluation results, we organize the figure by categorizing
the applications into three groups based on memory pressure:
(1) low memory pressure, (2) medium memory pressure, and (3) high memory pressure,
as shown in the figure. For applications---e.g., {\tt exchange2} and {\tt
imagick}---with low memory pressure, whose working sets are small enough to
fit in CPU's private caches \cite{singh2019memory}, both DDR5-R and CMM-H cause
negligible performance loss, as their memory accesses hit the CPU caches most
of time. On the other hand, applications like {\tt parest} and {\tt xalancbmk},
which have slightly larger working sets than the LLC size (256 MB), exhibit
moderate memory pressure and hence run slightly ($\approx$ 5\%) slower on DDR5-R.
A similar trend is observed for CMM-H, although these applications
experience greater slowdowns due to its higher latency than DDR5-R.

Nevertheless, as presented in the figure, applications like {\tt mcf} and {\tt
bwaves} experience significant performance losses on both DDR5 and CMM-H.
For example, {\tt bwaves} suffers a 1.37x and 3.73x slowdown on DDR5-R and CMM-H,
respectively. This is because these applications have large working sets, 
leading to high LLC misses, e.g., {\tt bwaves} and {\tt fotonick} suffer a 44\%
and 46\% LLC miss rate on our test server, respectively, causing many of their memory
accesses to
be served by the slower CMM-H. Among these, {\tt
fotonick} gets degraded dramatically because of its irregular memory accesses,
causing a 99.988\% DRAM cache hit rate. Note that such a high DRAM cache hit rate
still leads to a significant performance drop due to the wide speed gap between the
DRAM cache and the backing NAND flash; see Section \ref{sec:xsbench} for detailed
analyses.



\begin{mdframed}[style=MyFrame]
    \textit{Key takeaways: Applications with smaller memory footprints can be fully
    supported on CMM-H alone, similar to DDR5-R, without any noticeable performance
    drops. However, other applications will require more intelligent solutions,
    such as OS/VM enabled automatic memory tiering, to avoid noticeable losses.}
\end{mdframed}

\subsection{LLaMA 3}


LLaMA 3 \cite{llama3}, developed by Meta, marks a significant step forward in
open-source large language model (LLM) technology. Building on the foundation
established by LLaMA 2, this version offers models with various numbers of
parameters---e.g., 1 billion (1B), 8 billion (8B), and 70 billion (70B)
parameters---available in both pre-trained and instruction-tuned configurations.
To stress the memory system while keeping experiments manageable, we choose 1B
model from LLaMA 3.2 \cite{llama3.2_1b_model} with a memory consumption of 4740
MB and an up to 97\% LLC miss rate.

\begin{figure}[h!]
    \centering
    \includegraphics[width=1\linewidth]{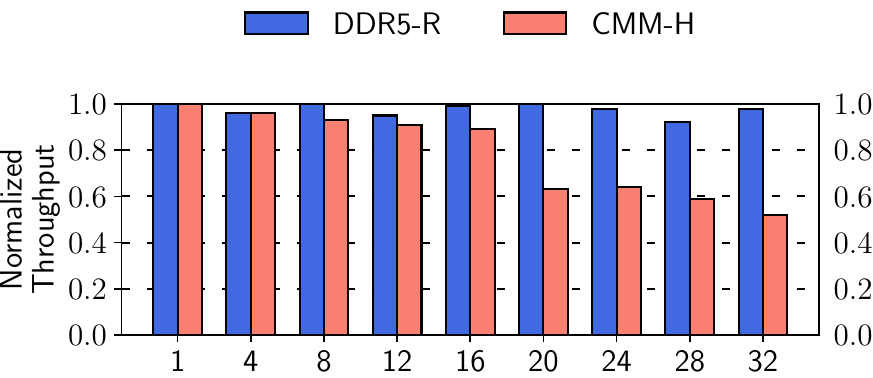}
    \caption{Normalized throughput (tokens/second) of LLaMA 3 with varying threads
    from 1 to 32 on DDR5-R and CMM-H to that on DDR5-L; higher is better}
    \label{fig:llama3_throughput}
\end{figure}

From Figure \ref{fig:llama3_throughput}, we can make two conclusions. First,
memory latency has an insignificant impact on the throughput of LLaMA 3. This
is because LLaMA 3's throughput remains stable across thread counts on DDR5-R,
although it is slower than DDR5-L by $\approx$ 50\% (see Figure
\ref{fig:micro_rand_lat} for details). Second, LLaMA 3's throughput is highly
affected by memory bandwidth. As shown in the figure, LLaMA 3's performance
loss gets larger as more threads run on CMM-H. For example, LLaMA 3 suffers a 4\% and
48\% slowdown in throughput with 4 and 32 threads, respectively. The reason
is that CMM-H's bandwidth saturates at four threads (see
Figure \ref{fig:cmmh_bw}) and more threads lead to higher bandwidth contention
on the CMM-H prototype.

\begin{mdframed}[style=MyFrame]
    \textit{Key takeaways: Bandwidth-sensitive applications can benefit most
    from CMM-H memory expansion, if the applications are intelligent enough in
    consolidating duplicate requests and if the CMM-H architecture, including the
    backing NAND flash, is optimized for maximum bandwidth utilization.}
\end{mdframed}

\subsection{XSBench}\label{sec:xsbench}


XSBench \cite{tramm2014xsbench} is widely used in high-performance computing (HPC)
research for performance evaluation and optimization of memory and compute-intensive
tasks. XSBench mimics the irregular memory access patterns and computational intensity
of full-scale Monte Carlo transport simulations, making it particularly valuable for
studying the behavior of memory hierarchies and bandwidth under irregular access
patterns.

In particular, XSBench offers users a way to configure problem sizes, such as the
number of energy levels, isotopes, and lookups, therefore allowing users to adjust
its memory footprint on demand; XSBench's memory consumption is proportional to
the problem size. By
leveraging this flexible configuration, it becomes possible for us to analyze how
the DRAM cache within CMM-H performs across varying memory footprints.
To achieve this, we select 10 different problem size configurations to
stress CMM-H, especially for its internal DRAM cache. For example, as shown in
Table \ref{tab:xsbench_config}, as the number of gridpoints per nuclide (specified
by flag -g) increases from 160 to 81920, XSBench's memory footprint starts from
only 80 MB and increases up to 40 GB, which enables different levels of the memory
hierarchy to be stressed. We run the single-threaded version of XSBench
(specified by flag -t 1) to avoid bandwidth contention on the CMM-H.

\begin{figure}[h!]
    \centering
    \includegraphics[width=1.0\linewidth]{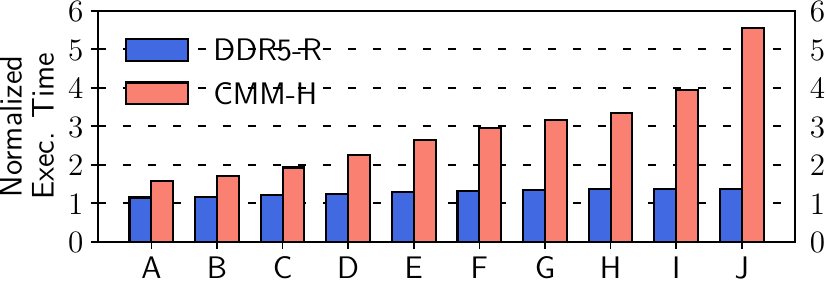}
    \caption{Normalized execution times of XSBench on DDR5-R and CMM-H, relative to those
    on DDR5-L; lower is better; x-axis stands for differing problem sizes
    and hence varying memory footprints (see Table \ref{tab:xsbench_config}
    for details)}
    \label{fig:xsbench}
\end{figure}

Figure \ref{fig:xsbench} shows XSBench's normalized execution times on DDR5 and
CMM-H, relative to those on DDR5-L, across the 10 configurations. As expected, 
the performance degradation of DDR5-R increases gradually with larger memory footprints. 
We attribute this stable performance degradation to the constant
inter-socket latency which is paid to access DDR5-R.

\begin{table}[h!]
    \centering
    \caption{XSBench Configurations}
    \resizebox{1.0\linewidth}{!}{
    \begin{tabular}{c||c|c|c}
        \hline
        Configuration & Data Input & Memory Footprint & \makecell{DRAM Cache\\Hit Rate} \\
        \hline
        A & -t 1 -g 160 & 80 MB & 99.9679\% \\
        \hline
        B & -t 1 -g 320 & 160 MB & 99.965\% \\
        \hline
        C & -t 1 -g 640 & 320 MB & 99.9624\% \\
        \hline
        D & -t 1 -g 1280 & 640 MB & 99.9502\% \\
        \hline
        E & -t 1 -g 2560 & 1279 MB & 99.9229\% \\
        \hline
        F & -t 1 -g 5120 & 2558 MB & 99.8635\% \\
        \hline
        G & -t 1 -g 10240 & 5117 MB & 99.8055\% \\
        \hline
        H & -t 1 -g 20480 & 10234 MB & 99.7653\% \\
        \hline
        I & -t 1 -g 40960 & 20468 MB & 98.6661\% \\
        \hline
        J & -t 1 -g 81920 & 40936 MB & 97.6209\% \\
        \hline
    \end{tabular}}
    \label{tab:xsbench_config}
\end{table}

In contrast, the performance loss for CMM-H increases rapidly. For example,
XSBench experiences a 1.58x performance loss when its memory footprint is
only 80 MB (configuration A in Table \ref{tab:xsbench_config}). This performance
loss goes up to 2.26x once the LLC cannot accommodate a memory footprint
of 640 MB (configuration D in the table). As expected, performance loss
remains the same level when the device DRAM cache can accommodate the memory
footprint, e.g., under configurations F, G, and H, where the DRAM cache
hit rate is consistently high as shown in the last column of Table
\ref{tab:xsbench_config}. Once the memory footprint becomes larger than
the device DRAM cache size (16 GB), the performance loss increases up to 3.94x
and 5.56x for configurations I and J, respectively. This is due to extensive
accesses to the significantly slow backing NAND flash. As shown in the last
column of Table \ref{tab:xsbench_config}, the DRAM cache hit rate drops to
only 97.6209\% for configuration J. Such a hit rate indicates that CMM-H's
latency is tens of times longer than that of the DRAM cache, given
the significant ($\approx$ 500x) speed gap between the backing NAND flash
and the DRAM cache; a performance analysis report from Samsung\cite{pm3a_perf}
uncovers that the random read and write latencies of the backend NAND flash
are on the order of tens of microseconds.

\begin{mdframed}[style=MyFrame]
    \textit{Key takeaways: Applications with irregular memory accesses
    and larger memory footprints than the DRAM cache capacity should
    expect to experience higher performance drops due to longer access
    latencies of the underlying NAND flash.}
\end{mdframed}

\section{Characterizing CMM-H as Memory Expander}

Note that CMM-H is not intended as a replacement for DRAM technology,
which would otherwise result in significant performance degradation
due to its longer latency and lower bandwidth than DRAM. Instead, CMM-H is
recommended to be used as a memory expansion solution to lower total cost of
ownership (TCO). As
shown in Table \ref{tab:mem_cost}, CMM-H achieves almost the same cost
as conventional SSDs, making it 9.2x and 24x more cost-effective than DDR4 and DDR5, respectively.

\begin{table}[h!]
    \centering
    \caption{Comparison of different memory devices}
    \begin{tabular}{c||c|c|c}
         \hline
         Memory Device & \$/GB & Persistency & Byte-Addressable \\
         \hline
         DDR4-2666 & 2.13 & No & Yes \\
         \hline
         DDR5-4800 & 5.62 & No & Yes  \\
         \hline
         SSD & 0.20 & Yes & No \\
         \hline
         CMM-H & 0.23 & Yes & Yes \\
         \hline
    \end{tabular}
    \label{tab:mem_cost}
\end{table}

In addition, given that the performance of CMM-H is on par with DRAM, it is
feasible to allocate part of program data to CMM-H without incurring significant 
performance degradation and at the same time enjoying lower TCO. To demonstrate
this, we choose Redis with two typical operations, i.e., GET and SET, to test
how their performance is affected by the
amount of program data allocated to CMM-H. For this purpose, we switch Linux
kernel to v6.12.12 such that we can use {\tt numactl} with {\tt
-weighted-interleave} flag to specify the ratios of memory pages allocated to
DDR5 and CMM-H. Table \ref{tab:redis_cmm_h} presents the normalized
throughputs of Redis GET and SET operations under varying DDR5-to-CMM-H page
allocation ratios compared to running the applications solely on DDR5-L. As shown
in the table, CMM-H achieves ~80\% of DDR5-L’s performance when 50\% of the
pages are allocated to the DDR5-L.
\begin{table}[h!]
    \centering
    \caption{Normalized throughputs of Redis operations with part of program
    data allocated to CMM-H to those of running the operations only on DDR5-L}
    \begin{tabular}{c||c|c|c}
         \hline
         Redis Operation & \makecell{75\% DDR5-L:\\25\% CMM-H} & \makecell{67\% DDR5-L:\\33\% CMM-H} & \makecell{50\% DDR5-L:\\50\% CMM-H} \\
         \hline
         GET & 91\% & 88\% & 84\% \\
         \hline
         SET & 87\% & 82\% & 74\% \\
         \hline
    \end{tabular}
    \label{tab:redis_cmm_h}
\end{table}

As a result, using CMM-H as a memory expansion is expected to 
yield significant TCO savings, especially for applications that do not always
require large memory space for peak performance.


\begin{mdframed}[style=MyFrame]
    \textit{Key takeaways: CMM-H is ready to be used as a memory expander for those
    applications that do not always require large memory space for peak performance
    while incurring acceptable performance degradation and gaining significant TCO
    savings.}
\end{mdframed}

\vspace{-5pt}
\section{Characterizing CMM-H as Persistent Memory}\label{sec:persist_mem}

With CXL GPF \cite{das2024introduction}, CMM-H works like
persistent memory, i.e., it survives all volatile data in the DRAM cache from
power failure. To show off how this feature enables higher performance for
persistent program, in this section, we evaluate two popular in-memory databases,
i.e., RocksDB \cite{rocksdb9.9.3} and Redis \cite{redis7.4.1}, which maintain
key-value stores in memory for fast data operations.

As described in Section \ref{sec:idem}, 
GPF alone cannot guarantee crash consistency. 
To address this issue, as with prior techniques
\cite{zeng2024compiler,liu2018ido,kim2020compiler}, we use
idempotent processing \cite{de2011idempotent} to partition input program into
a series of idempotent
and thus recoverable regions; please refer to Section \ref{sec:idem} for more
details. Following a prior work \cite{zeng2024compiler}, we implement idempotent
processing on top of Clang/LLVM 13.0.1 \cite{lattner2004llvm} to build RocksDB
9.9.3 \cite{rocksdb9.9.3} and Redis 7.4.1 \cite{redis7.4.1} from source code.
To ensure recoverability
for both workloads, we use the same compiler to build all the necessary
runtime libraries, e.g., {\tt glibc v2.27} \cite{glibc}, LLVM C++ library {\tt
libcxx} \cite{libcxx}, LLVM {\tt compiler-rt} \cite{compilerrt}, and LLVM stack
unwinding library {\tt libunwind} \cite{libunwind}, and link the binaries of
the RocksDB and Redis against those libraries.

It is worth noting that we pick idempotent processing among various crash 
consistency solutions due to its ability to transparently and automatically
delineate recoverable regions. In contrast, other techniques, e.g., Intel
PMDK \cite{pmdk}, Atlas \cite{chakrabarti2014atlas}, and Clobber-NVM
\cite{xu2021clobber}, require rewriting user source code with complex persistency
models \cite{pelley2014memory,kolli2017language} in mind, posing significant
programming burden on users and potentially generating program bugs \cite{liu2019pmtest,neal2021hippocrates,liu2020cross,di2021fast}.

\subsection{RocksDB}

RocksDB \cite{rocksdb9.9.3} is a high-performance key-value store developed by
Facebook. Its core structure is an in-memory table called memtable which is
technically a sorted data structure to accomplish efficient insertions and fast
searches.
When users issue a write
request, the data being stored first goes into the memtable. To guarantee crash
persistence in case of using DRAM as main memory, RocksDB also appends
the data to a write-ahead log (WAL) file---residing in external storage---for
failure recovery along with storing the data in the memtable, which however is
considerably expensive.

\ignore{
To optimize performance, RocksDB ensures all I/O operations are sequential. That
way, it can take advantage of the sequential access strengths of hard disks. In
addition, it allows concurrent writes even when the memtable is being flushed to
disk---the memtable is flushed to hard disks when it is full---and predominantly
performs large writes, with the exception of WAL appends. However, WAL appends
could negatively impact performance on bandwidth-limited memory systems, e.g.,
Samsung CMM-H and Intel PMEM.
}

\begin{figure}[h!]
    \centering
    \includegraphics[width=1.0\linewidth]{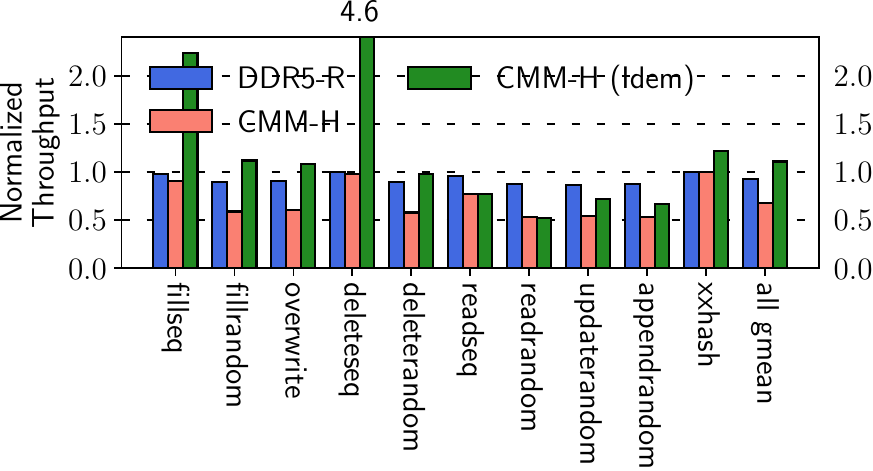}
    \caption{Normalized throughputs of RocksDB on DDR5-R, CMM-H, and CMM-H (idem)
    to those on DDR5-L; higher is better; x-axis stands for RocksDB's benchmarks}
    \label{fig:rocksdb}
\end{figure}

To demonstrate how CMM-H benefits RocksDB by eliminating the expensive WAL, we
run {\tt db\_bench} \cite{cao2020characterizing} for four configurations,
DDR5-L, DDR5-R, CMM-H, and
CMM-H (idem). We enable WAL for the original RocksDB running on the former three
memory devices, while disabling WAL for {\it idempotence-processed RocksDB} on
the CMM-H. Figure \ref{fig:rocksdb} presents how RocksDB performs for these
configurations. The x-axis indicates benchmarks from {\tt db\_bench}, while
the y-axis stands for the normalized throughputs of these benchmarks to those
on DDR5-L (baseline).
\textbf{\textit{Note that CMM-H (idem) stands for the normalized throughputs of
the benchmarks compiled by our idempotent compiler on the CMM-H; see Section
\ref{sec:idem} for the explanation of idempotent processing.}}

It is not surprising to see from the figure, that CMM-H leads to a higher performance
slowdown than DDR5-L and DDR5-R. Due to longer latency and lower bandwidth, CMM-H
(red bar in the figure) leads an average of 27\% performance degradation, which
is higher than only 7\% of DDR5-R. In addition, we can see from the figure, CMM-H
results in a significant performance loss (41\%) for {\tt fillrandom} which exhibits
heavy random writes. This is because, due to randomness, {\tt fillrandom} experiences
high amount of misses in the DRAM cache. As such, a significant overhead is paid
to access the backing NAND flash of the CMM-H prototype.

Nevertheless, DDR5-R and CMM-H lead to a marginal performance loss for benchmarks
with heavy sequential writes. For example, there is no performance degradation
observed for {\tt fillseq} and {\tt deleteseq} on DDR5-R as writes are usually not
on the critical path. CMM-H results in an acceptable (i.e., 9\%) slowdown for
both benchmarks due to longer write latency (see Figure \ref{fig:cmmh_bw}).

\begin{mdframed}[style=MyFrame]
    \textit{Key takeaways: Users should avoid performing random writes to CMM-H and
    instead employ intelligent software to coalesce them at page (4 kB) granularity.
    As such, the applications can benefit from page-level locality in the DRAM
    cache of the CMM-H prototype.}
\end{mdframed}


As reads are on the critical path of the core pipeline execution, a longer memory
access latency could result in a severe performance loss. The figure shows that
DDR5-R exhibits a 4\% and 12\% slowdown for sequential read ({\tt readseq}) and
random read ({\tt readrandom}), respectively. Of course, due to higher latency
than DDR5, CMM-H leads to even higher performance degradations for those two
operations; higher randomness results in poor data locality, which renders the
DRAM cache ineffective. As such, the figure shows that {\tt readseq} and {\tt
readrandom} observe a 23\% and 47\% performance loss, respectively.


Interestingly, despite being slower than DDR5-L and DDR5-R, CMM-H brings higher
throughputs for write-intensive applications when it is enabled as persistent memory
and the applications are idempotent.
The figure shows that CMM-H boosts the throughputs of those evaluated programs
by 11\% on average; especially for {\tt fillseq} and {\tt deleteseq}, they witness
a 2.2x and 4.59x speedup, respectively. The reason is twofold: (1) the slow CMM-H
has marginal impact on those applications as they heavily use stores which are not
on the critical path; (2) with idempotent processing and CXL GPF \cite{fridman2023cxl},
those applications are crash-consistent, thereby obviating the need to use expensive
write-ahead log (WAL). 

\begin{mdframed}[style=MyFrame]
    \textit{Key takeaways: By obviating the need for expensive WAL, CMM-H with
    idempotent processing offers a chance to achieve significant performance gains
    for write-heavy applications, compared to the reliance on WAL to ensure crash 
    consistency.}
\end{mdframed}

\subsection{Redis}

Redis \cite{redis7.4.1} is another high-throughput in-memory key-value store commonly employed
in website development as a caching layer and for message queuing applications. To ensure
crash recovery, Redis logs all write operations to an append-only file (AOF), which is
managed in DRAM main memory and
periodically flushed to external storage devices based on a predefined flushing policy.
Redis allows users to configure the frequency of AOF flushing to balance performance and
consistency: higher flushing frequency improves recoverability but reduces performance
due to increased interactions with external storage devices.

Specifically, Redis supports three AOF flushing modes: {\tt always}, {\tt everysec},
and {\tt no}. In our experiments, we apply the default {\tt always} flushing policy for
DDR5-L, DDR5-R, and CMM-H (it is configured as volatile memory). This policy ensures
that the AOF is flushed to external persistent devices after every log append,
guaranteeing no data loss in the event of a crash. Conversely, for CMM-H (idem), we
disable the logging and use the {\tt no} flushing policy to eliminate the overhead of the
logging. Instead, crash consistency is maintained through GPF \cite{fridman2023cxl}
and idempotent processing, which work together to ensure data consistency without
logging. By employing CMM-H (idem), we can estimate the potential performance gains
that CMM-H can achieve for Redis.

\begin{figure}[h!]
    \centering
    \includegraphics[width=1.0\linewidth]{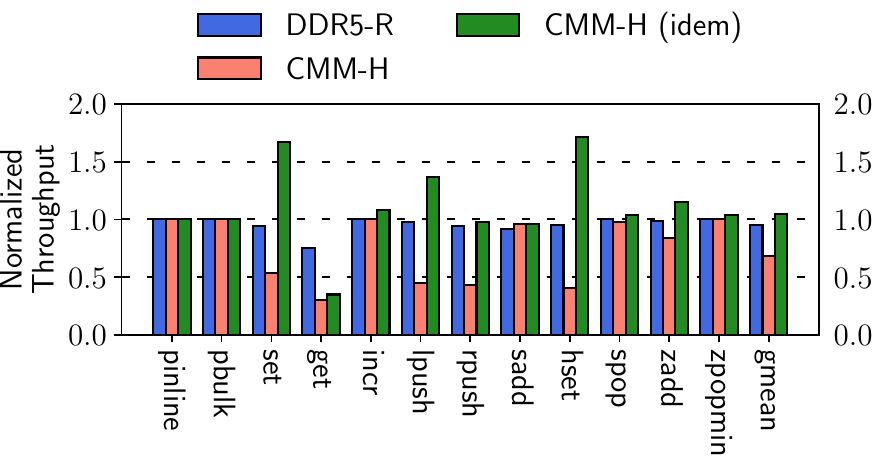}
    \caption{Normalized throughputs of varying Redis applications on DDR5-R, CMM-H,
    and CMM-H (idem) to those on DDR5-L; higher is better; x-axis shows the applications
    from redis-benchmark}
    \label{fig:redis}
\end{figure}

Figure \ref{fig:redis} shows the normalized throughputs of 12 applications for three
memory configurations: DDR5-R, CMM-H, and CMM-H (idem). In the figure, the bars of DDR5-R
and CMM-H represent the normalized throughputs of original Redis applications
running on those two memory devices, respectively, while CMM-H (idem) stands for
the normalized throughputs of the {\it idempotent-processed} Redis applications
with logging disabled on CMM-H.

As shown in the figure, CMM-H incurs a performance loss for
memory-intensive applications, e.g., 46\% for {\tt set} and 70\% for {\tt
get}. On the other hand, CMM-H leads to no observable negative performance
impact on some applications, e.g., {\tt pinline (ping\_inline)}, {\tt pbulk
(ping\_bulk)}, and {\tt sadd}. We suspect this is because their memory accesses
inherently hit in CPU caches, rendering those applications not sensitive to memory
speed and bandwidth.

\begin{mdframed}[style=MyFrame]
    \textit{Key takeaways: Despite being slower than DRAM, CMM-H is a good fit for
    compute-intensive applications that exhibit high data locality and hence
    generate less frequent data accesses to main memory.}
\end{mdframed}

In particular, we can see from the figure, that with idempotent processing and GPF
technique, CMM-H (idem) instead drastically improves the throughputs of some
applications, such as 1.67x for {\tt set}, 1.37x for {\tt lpush}, and 1.71x for
{\tt hset}. This is because CMM-H (idem) could offset the lower performance of
CMM-H by eliminating the expensive logging to external persistent devices. Even
though CMM-H (idem) does not improve throughputs of other applications a lot,
e.g., {\tt rpush}, {\tt spop}, and {\tt zpopmin}, it offers these applications
crash consistency and obviating the expensive logging, leading to shorter tail
latency and ultimately improving user experience.

\begin{mdframed}[style=MyFrame]
    \textit{Key takeaways: As a replacement for the combination of DRAM and slower
    storage devices, CMM-H has the potential to be deployed in production environments
    for persistent applications to significantly boost their performance.}
\end{mdframed}

\section{Other Related Work}

Compute Express Link (CXL) has emerged as a key enabling technology for memory expansion and pooling in modern datacenters. Extensive research has examined the potential of CXL-based memory for disaggregated architectures \cite{gouk2023memory, jang2023cxl, li2023pond, gouk2022direct, aguilera2023memory, wang2023cxl,giannoula2023daemon}. Other research has investigated the role of CXL in tiered memory systems \cite{maruf2023tpp, yuhong2024managing, zhou2024neomem, kim2023smt, sha2023vtmm}. Notably, Meta has investigated using CXL for memory tiering \cite{maruf2023tpp, weiner2022tmo}, while Microsoft has explored using CXL for memory disaggregation \cite{berger2023design,li2023pond} in production environments. 

However, due to the limited availability of commercial CXL hardware, most studies rely on simulations and emulations for performance analysis \cite{yang2023cxlmemsim, wang2024asynchronous, esmaili2024mess, fridman2023cxl, arif2022exploiting}. While these studies provide valuable insights, they cannot fully capture the complexities of real-world deployments. To bridge this gap, several projects developed CXL hardware prototypes \cite{gouk2022direct,maruf2023tpp}, facilitating empirical evaluations and expanding our understanding of CXL's potential in practical settings.

Nonvolatile memory (NVM) technologies, such as ReRAM \cite{akinaga2010resistive, chen2020reram}, PCM \cite{kim2019ll, qureshi2009enhancing, sengupta2015framework, tyagi2007pcm}, STT-MRAM \cite{chi2016architecture, huai2008spin, khvalkovskiy2013basic, korgaonkar2018density}, 3D XPoint \cite{hady2017platform}, Intel Optane memory (PMEM) \cite{intel2023pm, shanbhag2020large, peng2019system, IntelPMmode, zardoshti2020understanding}, offer viable alternatives to DRAM. In this context, CXL has opened new opportunities to advance the performance of persistent memory and its adoption in modern computing systems \cite{kwon2023failure, fridman2023cxl, samsung2024cmmh}. TrainingCXL \cite{kwon2023failure} is such an example that leverages CXL-enabled disaggregated memory to integrate persistent memory and GPUs within a cache-coherent domain. Additionally, Fridman et al. \cite{fridman2023cxl} explored the viability of CXL as a persistent memory solution for disaggregated high-performance computing, demonstrating its potential using an FPGA prototype with memory-backed DRAM. Samsung’s CMM-H prototype further exemplifies these advancements by combining a DRAM cache with NAND flash in a single device.
Despite the innovations of CMM-H, the limited availability of publicly available performance characterizations has hindered its widespread adoption and the development of optimized programming practices. To address this gap, this paper presents the first comprehensive performance evaluation of the CMM-H prototype. 





\section{Conclusion}
This paper provides the first comprehensive performance evaluation of Samsung’s FPGA-based CMM-H prototype, which provides an economical way to significantly expand memory capacity while optionally offering persistence, compared to conventional memory technologies. To achieve high performance, CMM-H utilizes a DRAM cache to buffer hot data of its backend NAND flash. With extensive experiments and data analyses, this paper uncovers that CMM-H exhibits longer random memory access latency and limited bandwidth compared to both local DRAM and remote DRAM. Despite the longer random latency and lower bandwidth, CMM-H is able to deliver comparable performance for many popular applications. Nevertheless, applications with irregular memory access patterns and large memory footprints should account for the finite DRAM cache, as exceeding its capacity can lead to significant performance degradation.

It is worth noting that CMM-H's advantages become prominent for persistent applications as it supports Global Persistent Flush (GPF) to flush the DRAM cache contents to the NAND flash upon power failure. For instance, CMM-H yields up to 4.6x and 1.7x throughput improvements for RocksDB and Redis, respectively. These performance gains primarily stem from the elimination of expensive logging to external storage devices by CMM-H and idempotent processing. Based on these experimental results, this paper offers key insights into how to best use the CMM-H device for both volatile and persistent applications.



\balance
\bibliographystyle{IEEEtranS}
\bibliography{refs}

\end{document}